\documentclass[journal=nalefd,manuscript=letter,layout=twocolumn]{achemso}

\usepackage{graphicx}
\usepackage[hidelinks]{hyperref}
\usepackage{amsmath,amssymb}
\usepackage{siunitx}

\usepackage[normalem]{ ulem }

\setkeys{acs}{usetitle=true}

\newcommand{\AffIMN}{Nantes Universit\'e, CNRS, Institut des Mat\'eriaux de Nantes Jean Rouxel (IMN), Nantes, France}
\newcommand{\AffIPR}{Univ Rennes, CNRS, Institut de Physique de Rennes (IPR) -- UMR 6251, 35000 Rennes, France}
\newcommand{\AffESRF}{ESRF, The European Synchrotron, 71 Avenue des Martyrs, CS40220, 38043 Grenoble Cedex 9, France}
\newcommand{\AffFEMTOMAX}{MAX IV Laboratory, Lund University, Lund, Sweden}

\title{Anisotropy of Ultrafast Strain in V$_2$O$_3$ Thin Films: Out-of-Equilibrium Phase Transitions under Interfacial Clamping}

\author{J. Guzman-Brambila}
\email{Julio-Cesar.GUZMAN-BRAMBILA@cnrs-imn.fr}
\affiliation{\AffIMN}
\affiliation{\AffIPR}

\author{R. Mandal}
\affiliation{\AffIPR}
\affiliation{\AffESRF}

\author{D. Lea}
\affiliation{\AffIPR}

\author{E. Trzop}
\affiliation{\AffIPR}

\author{M. Servol}
\affiliation{\AffIPR}

\author{J. C. Ekstr\"om}
\affiliation{\AffFEMTOMAX}

\author{F. Pawula}
\affiliation{\AffIMN}

\author{J. Tranchant}
\affiliation{\AffIMN}

\author{L. Cario}
\affiliation{\AffIMN}

\author{M. Lorenc}
\affiliation{\AffIPR}

\author{E. Janod}
\email{etienne.janod@cnrs-imn.fr}
\affiliation{\AffIMN}

\author{C. Mariette}
\affiliation{\AffESRF}
\email{celine.mariette@esrf.fr}

\abbreviations{XRD,TR-XRD,AFI,IMT}

\let\oldmaketitle\maketitle
\let\maketitle\relax

\begin{document}


\twocolumn[
\begin{@twocolumnfalse}
\oldmaketitle
\begin{abstract}

Ultrafast photoinduced insulator-to-metal transitions in correlated materials are often mediated by lattice distortions, yet the role of interfacial lattice constraints in shaping nonequilibrium pathways remains largely unexplored. We use azimuth-resolved time-resolved X-ray diffraction to track orientation-dependent strain dynamics in granular V$_2$O$_3$ thin films on c-cut sapphire, where thermal-expansion mismatch imposes anisotropic interfacial strain. Across the thermal transition, the azimuthal profile of the (110)$_H$ strain inverts curvature, providing direct evidence of partial clamping of the hexagonal basal-plane lattice (a$_H$,b$_H$). After photoexcitation of the antiferromagnetic insulating phase, the structural response remains clamp-limited: weakly constrained grain families reach the full basal-plane contraction characteristic of the metallic-like state, whereas strongly constrained families exhibit a strongly reduced distortion. Fluence-dependent measurements further disentangle transformed fraction from clamping-limited lattice distortion. Our results show that interfacial clamping acts as a static selector for ultrafast phase switching and provide a general route to quantify anisotropic strain dynamics in heterostructures.
\end{abstract}
\end{@twocolumnfalse}
]


\section{Introduction}

Ultrafast control of electronic phases in correlated materials is a central objective in condensed matter physics, motivated by prospects for high-speed switching and functional devices\cite{basovPropertiesDemandQuantum2017}. In strongly correlated systems, changes of electronic or magnetic state couple strongly to the lattice, so optical electronic excitation can in general trigger structural dynamics \cite{giannettiUltrafastOpticalSpectroscopy2016}. When ultrashort and intense laser pulses are applied, these structural changes can be ultrafast and reshape the electronic landscape. A key open question is how such nonequilibrium pathways are set in thin films and heterostructures, where the lattice is not free but constrained by an interface\cite{abadiasReviewArticleStress2018, murakamiThermalStrainLead1977, wiederCalculationThermallyInduced1995}.

Recent time-resolved work has established that, in V$_2$O$_3$, photoexcitation generates strain waves that actively participates in driving the insulator-to-metal transition, emphasizing that lattice distortions are not merely a byproduct but part of the switching mechanism\cite{amanoPropagationInsulatortometalTransition2024, lantzUltrafastEvolutionTransient2017, singerNonequilibriumPhasePrecursors2018, ronchiEarlystageDynamicsMetallic2019}. In thin films, however, strain is also strongly affected by interfacial stress imposed by the substrate. The latter may result from growth conditions and/or thermal-expansion mismatch between film and substrate, accumulated upon cooling from growth or annealing\cite{murakamiThermalStrainLead1977, wiederCalculationThermallyInduced1995, sakaiTransportPropertiesRatio2015, barazaniPositiveNegativePressure2023}. The effect of interfacial strain in static conditions has been broadly studied, and interfacial strain is nowadays seen as a tunable parameter, either to avoid related limitations or to exploit related opportunities\cite{hommRoomTemperatureMott2021, sakaiTransportPropertiesRatio2015, barazaniPositiveNegativePressure2023, liInsightsStrainEngineering2024, damodaranNewModalitiesStraincontrol2016}.

Here we show that interfacial clamping acts as a static selector for the photoinduced structural pathway in textured polycrystalline V$_2$O$_3$ films deposited on sapphire. Using azimuth- and time-resolved X-ray diffraction, we isolate grain families by orientation and directly measure the anisotropic response of the hexagonal basal-plane lattice (a$_H$,b$_H$) via the (110)$_H$ reflection. Across the thermal transition at $T_{PM-AFI}$$\simeq150$ K, the azimuthal strain profile changes curvature, providing a clear signature of partial basal-plane clamping: grain families most strongly coupled to the interface exhibit a markedly reduced lattice distortion compared to weakly clamped orientations. Crucially, this constraint persists out of equilibrium: following photoexcitation of the low-temperature AFI phase, the transient response remains strongly orientation dependent, with weakly clamped grains reaching the full basal-plane contraction characteristic of the metallic-like state, while strongly clamped grains display a substantially suppressed structural distortion.

By separating the fluence-dependent converted fraction from the orientation-dependent maximum lattice distortion, our results establish interfacial clamping as a deterministic boundary condition that selects the structural metric reached during ultrafast phase switching in V$_2$O$_3$ thin films. More broadly, the approach provides a general route to quantify and exploit anisotropic boundary conditions in ultrafast transitions of thin-film materials where structural degrees of freedom are central to the phase transformation\cite{abadiasReviewArticleStress2018, cancellieriStrainDepthProfiles2021, amanoPropagationInsulatortometalTransition2024, lantzUltrafastEvolutionTransient2017, singerNonequilibriumPhasePrecursors2018}.


\section{Methods}

\begin{figure*}[t]
\centering
\includegraphics[width=1.0\textwidth]{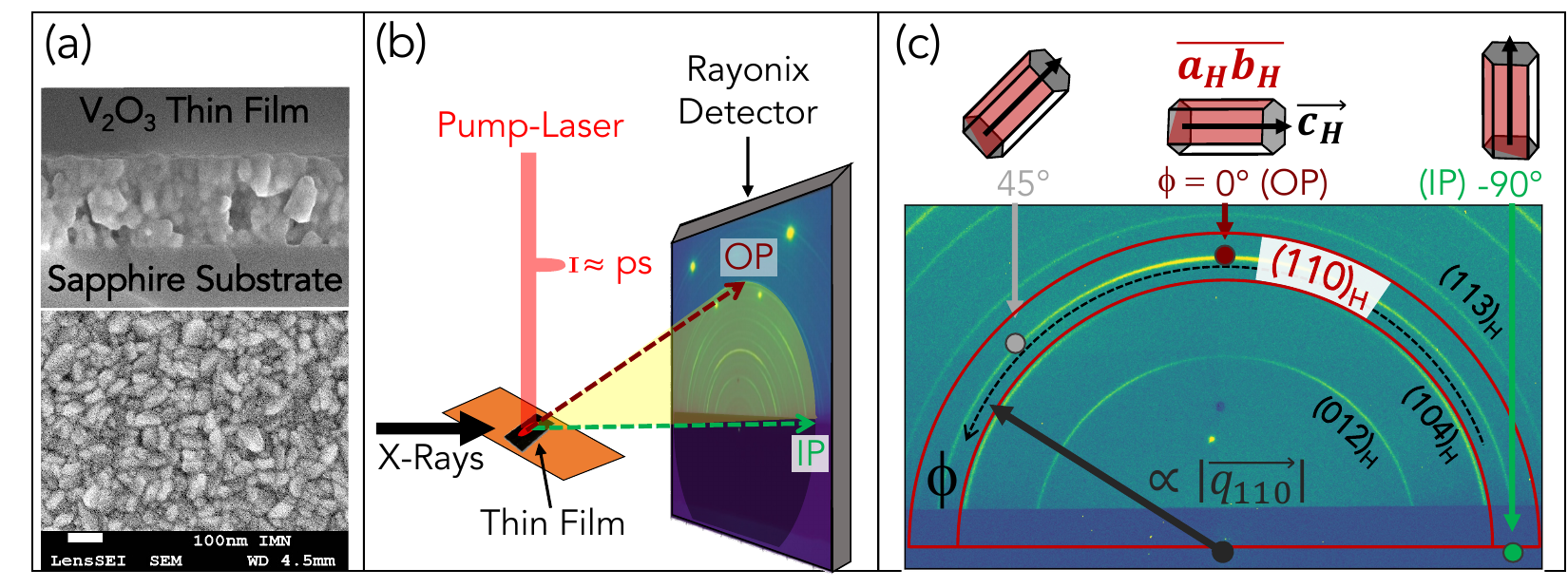}
\caption{\label{fig:fig1}
(a) Representative SEM images of the textured polycrystalline V$_2$O$_3$ thin films, shown in side view (top) and top view (bottom), illustrating the granular microstructure.
(b) Schematic of the pump-probe time-resolved X-ray diffraction experiment. The thin film is probed by a horizontal X-ray beam at a fixed incident angle, either $2^{\circ}$ or $6^{\circ}$, and photoexcited by a near-infrared pump pulse incident normal to the film surface. The green and dark-red arrows indicate the azimuthal directions associated with in-plane (IP) and out-of-plane (OP) orientations of the $(110)_{H}$ scattering vector with respect to the substrate plane.
(c) Representative two-dimensional diffraction pattern showing the hexagonal $(110)_{H}$ diffraction ring. The azimuthal angle $\Phi$, used for caking analysis, is indicated. Schematics illustrate the crystallite orientations contributing to diffraction at selected $\Phi$ values, relating the measured azimuthal position to the orientation of the $(a_{H},b_{H})$ basal plane and the $c_{H}$ axis with respect to the film/substrate interface.}
\end{figure*}

Textured polycrystalline (V$_{1-x}$Cr$_x$)$_2$O$_3$ and V$_2$O$_3$ thin films were deposited on c-cut $(0001)$ sapphire substrates by reactive magnetron co-sputtering from vanadium and chromium targets in a reactive Ar (99.25\%)/O$_2$ (0.75\%) plasma at a pressure of 25 $\mu$bar. Post-deposition annealing at 500 $^{\circ}$C under a reducing Ar (95\%)/H$_2$ (5\%) flow crystallized the films\cite{querreMetalInsulatorTransitions2016}. Representative scanning electron microscopy (SEM) images of the resulting granular microstructure are shown in Fig. \ref{fig:fig1}(a).

Static and time-resolved X-ray diffraction measurements were performed at ID09 (ESRF)\cite{wulffRealizationSubnanosecondPump2003, cammarataChopperSystemTime2009} using 15 keV photons selected by a Si(111) channel-cut monochromator with $\Delta E/E = 10^{-4}$. Diffraction images were recorded on a 2D Rayonix MX170-HS detector placed at a sample-detector distance of 120 mm. The experimental geometry is schematized in Fig. \ref{fig:fig1}(b). Room-temperature pump-probe measurements were performed at an X-ray incidence angle of $\approx 2^{\circ}$, which provided access to an extended azimuthal range. Temperature-dependent measurements and low-temperature pump-probe measurements were performed at an incidence angle of $6^{\circ}$, chosen as a compromise between the X-ray footprint on the sample, spatial resolution, and accessible azimuthal range. For low-temperature measurements, the samples were mounted on the cold finger of a continuous N$_2$-flow cryostat equipped with Mylar windows for X-ray access and a fused-silica window for optical transmission.

Time-resolved measurements were performed in a pump-probe scheme using optical excitation delivered by a 2 ps laser pulse (TOPAS, Light Conversion), synchronized to the $\approx$100 ps X-ray pulses. The X-ray pulse duration sets the time resolution for these measurements. Room-temperature pump-probe measurements were performed using 1.55 eV excitation, whereas low-temperature photoinduced AFI$\rightarrow$PM-like measurements were performed using 0.9 eV excitation. The laser arrived normal to the film surface and was shaped to match the elongated X-ray footprint. The typical laser spot size on the sample was 1.2 mm along the X-ray direction $\times$ 0.3 mm, much larger than the X-ray footprint (250 $\mu$m x 25 $\mu$m). Measurements were performed in stroboscopic mode with a 1 kHz repetition rate.

To resolve orientation-dependent lattice distortions, we applied an azimuthal caking procedure\cite{kiefferPyFAIVersatileLibrary2013, abadiasReviewArticleStress2018} by integrating the 2D diffraction intensity over the azimuthal angle $\Phi$, defined in Fig. \ref{fig:fig1}(c). The absence of artificial azimuthal shifts from the detector geometry and caking procedure was verified using a LaB$_6$ powder reference, as described in Section S1 of the Supporting Information. This procedure separates the $(110)_{H}$ response according to the orientation of the scattering vector with respect to the substrate plane, from out-of-plane (OP) sectors near $\Phi \approx 0^{\circ}$ to in-plane (IP) sectors toward $\Phi \rightarrow \pm 90^{\circ}$ (\ref{fig:fig1} b), within the accessible angular range. For room-temperature pump-probe measurements, the accessible caking range was $\Phi \in [-80^{\circ},+80^{\circ}]$. For temperature-dependent and low-temperature pump-probe measurements, the accessible range was limited to $\Phi \in [-60^{\circ},+60^{\circ}]$ due to the larger X-ray incidence angle. In both cases, the data was integrated using 20$^{\circ}$-wide azimuthal windows centered every 20$^{\circ}$.

The analysis focused on the hexagonal $(110)_{H}$ reflection. As depicted in Fig. \ref{fig:fig1}(c), the out-of-plane (OP) sectors mainly probe grain families with the $(a_{H},b_{H})$ basal plane approximately perpendicular to the substrate plane, whereas the In-Plane (IP) sectors probe grain families with the basal plane closer to parallel to the substrate surface. After azimuthal integration, the $(110)_{H}$ peak in each $\Phi$ slice was fitted with a pseudo-Voigt profile to extract the average scattering-vector position $q_{110}$, providing a direct measure of the basal-plane lattice response as a function of crystallite orientation. The same procedure was applied to static and time-resolved measurements.


\section{Results and Discussion}

\begin{figure*}[h!]
\centering
\includegraphics[width=1\textwidth]{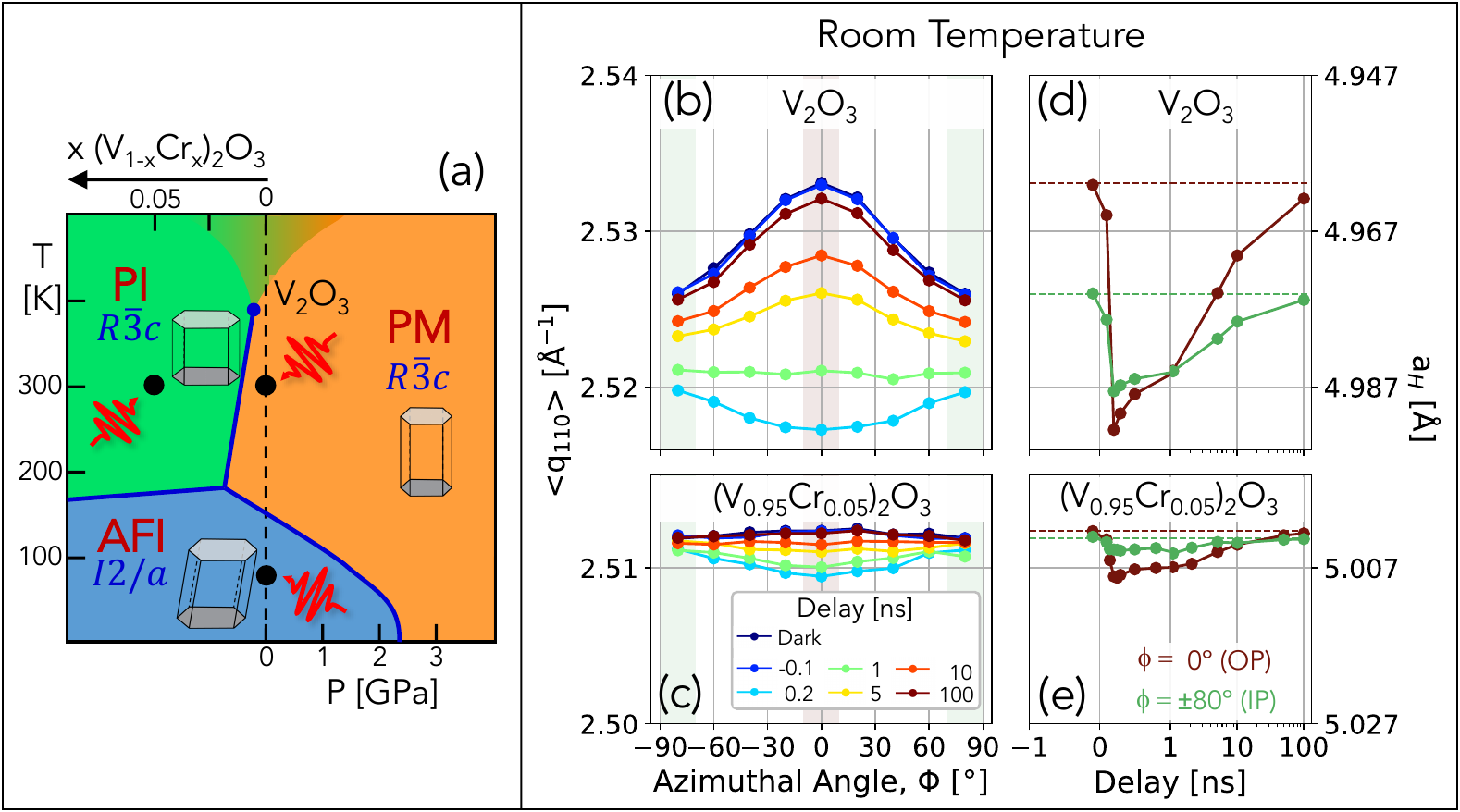}
\caption{\label{fig:fig2}
(a) Temperature--composition--pressure phase diagram of the (V$_{1-x}$Cr$_x$)$_2$O$_3$ system, highlighting the experimental conditions for the azimuth-resolved structural study performed in this work. These include room-temperature photoexcitation for $x = 0$ in the PM phase and $x = 0.05$ in the PI phase, temperature-dependent measurements for $x = 0$ across the PM-AFI transition, and low-temperature photoexcitation for $x = 0$ in the AFI phase. (b,c) Azimuth-resolved room-temperature pump-probe response of $\langle q_{110}\rangle$, equivalently the corresponding $a_H$ lattice parameter, for (b) V$_2$O$_3$ and (c) (V,Cr)$_2$O$_3$ with 5\% Cr. Data are extracted from 20$^{\circ}$-wide azimuthal sectors over $\Phi = -80^{\circ}$ to $+80^{\circ}$, before excitation (dark) and at different delays after photoexcitation. (d,e) Time evolution of $\langle q_{110}\rangle$ and the corresponding $a_H$ lattice parameter for $\Phi = 0^{\circ}$ and $|\Phi| = 80^{\circ}$. The pump photon energy was 1.55 eV, the pulse duration was 2 ps, and the pump fluence was 21 mJ/cm$^{2}$.}

\end{figure*}

We first focus on the hexagonal $(110)_{H}$ reflection, which directly probes the basal-plane metric of V$_2$O$_3$. This reflection is particularly relevant because the (a$_H$,b$_H$) lattice parameters act as a structural order parameter for the PM-AFI transition\cite{mcwhanMetalInsulatorTransitionV1xCrx2O31970} and are closely linked to the metallic or insulating character of the material\cite{hommRoomTemperatureMott2021}. In the following, the average scattering-vector position $\langle q_{110}\rangle$ is used to track the basal-plane response, while the azimuthal angle $\Phi$ separates grain families with different orientations, as described in Fig. \ref{fig:fig1}. As illustrated in Fig. \ref{fig:fig2}(a), we have monitored this grain orientation-dependent basal-plane response under both ultrafast photoexcitation in the three PI, PM and AFI domains of the  V$_2$O$_3$ phase diagram and temperature variations across the PM-AFI transition.

Figure \ref{fig:fig2}(b-e) show the room-temperature photoinduced response of $\langle q_{110}\rangle(\Phi)$ for pure V$_2$O$_3$ (PM phase) and for (V,Cr)$_2$O$_3$ with 5\% Cr (PI phase). Before excitation, both compounds already display an azimuth-dependent peak position, showing that the basal-plane metric depends on the crystallite orientation. For pure V$_2$O$_3$, this laser-off concave azimuthal profile is also observed in films with different thicknesses, indicating that the orientation-dependent strain is not strongly relaxed over the 100-300 nm thickness range investigated here, as shown in Section S2 of the Supporting Information. After photoexcitation, the $(110)_{H}$ feature shifts toward lower $q_{110}$, corresponding to an increase of the apparent $a_H$ lattice parameter. The distortion is not uniform over $\Phi$: for both compounds, the response is more pronounced near $\Phi \approx 0^{\circ}$ than at higher angle, indicating that the out-of-plane sectors deform more strongly than their in-plane counterparts.

A second important observation is that the curvature of $\langle q_{110}\rangle(\Phi)$ changes upon photoexcitation. Before excitation, the profile is concave, with larger $\langle q_{110}\rangle$ values near $\Phi \approx 0^{\circ}$ and smaller values toward large $|\Phi|$. At the delay of maximum response, the profile evolves toward the opposite curvature, because the deformation is larger near $\Phi \approx 0^{\circ}$ than in the large-$|\Phi|$ sectors. This suggests that grain families for which the (a$_H$,b$_H$) basal plane is approximately perpendicular to the substrate plane can accommodate the photoinduced deformation more freely than those for which the basal plane is closer to parallel to the substrate. At this stage, however, the room-temperature data alone do not establish the origin of this orientation dependence.

\begin{figure*}[h!]
\centering
\includegraphics[width=1.0\textwidth]{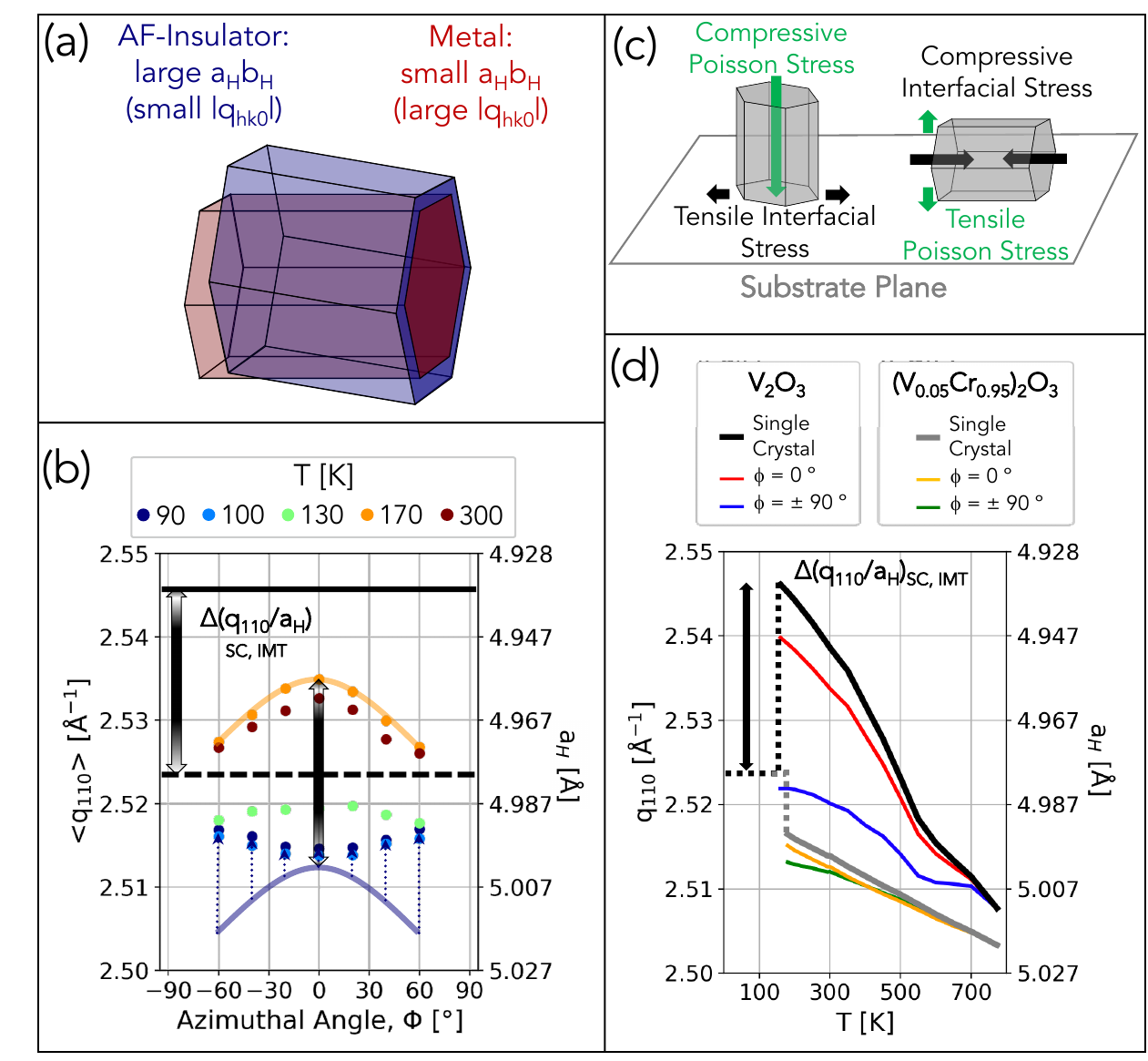}
\caption{\label{fig:fig3}
(a) Schematic representation of the metallic PM hexagonal unit cell (red) and the corresponding distorted hexagon in the low-symmetry monoclinic AFI insulating phase (blue).
(b) Azimuth-resolved $\langle q_{110}\rangle$, and corresponding $a_H$ lattice parameter, measured across the PM-AFI transition for 20$^{\circ}$-wide sectors between $\Phi = -60^{\circ}$ and $+60^{\circ}$. The orange curve is a guide to the eye for the PM profile at 170 K; the blue shaded curve shows the AFI profile expected for a $\Phi$-independent transition jump. Arrows highlight the suppression of the transition-induced basal-plane expansion at large $|\Phi|$.
(c) Schematic representation of the substrate-induced elastic coupling for two limiting crystallite orientations: $\Phi = \pm90^{\circ}$, where the (a$_H$,b$_H$) plane is parallel to the substrate plane, and $\Phi = 0^{\circ}$, where $c_H$ lies in the substrate plane. The sketches indicate how interfacial stress generates Poisson-coupled strain along the unconstrained crystallographic directions.
(d) Temperature dependence of $q_{110}$ and $a_H$ for V$_2$O$_3$ (black) and (V,Cr)$_2$O$_3$ with 5\% Cr (gray), adapted from McWhan et al. \cite{mcwhanMetalInsulatorTransitionV1xCrx2O31970}, together with the calculated evolution expected for films on c-cut sapphire for the two limiting orientations shown in panel (c): V$_2$O$_3$ in red and blue, and (V,Cr)$_2$O$_3$ with 5\% Cr in orange and green.}
\end{figure*}

To clarify this point, we next examine the temperature-driven PM-AFI transition of pure V$_2$O$_3$. As schematized in Fig. \ref{fig:fig3}(a), the high-temperature PM phase has a smaller basal-plane lattice parameter than the low-temperature AFI phase. The PM$\rightarrow$AFI transition is therefore associated with an expansion of the (a$_H$,b$_H$) plane, observed experimentally as a decrease of $\langle q_{110}\rangle$\cite{mcwhanMetalInsulatorTransitionV1xCrx2O31970}.

Figure \ref{fig:fig3}(b) reports $\langle q_{110}\rangle(\Phi)$ across the PM-AFI transition for azimuthal sectors between $\Phi = -60^{\circ}$ and $+60^{\circ}$. In the PM phase, at room temperature and at 170 K, the profile is concave, as already observed in the room-temperature pump-probe data. Upon cooling through the transition, $\langle q_{110}\rangle$ decreases, meaning that the basal plane expands. However, this expansion is strongly orientation dependent. The decrease in $\langle q_{110}\rangle$ is largest near $\Phi \approx 0^{\circ}$, while it is progressively reduced at larger $|\Phi|$. As a consequence, the azimuthal profile changes from concave in the PM phase to convex in the AFI phase.

Such an orientation-dependent suppression is consistent with previous reports showing that the strain state and the electronic transition behavior of V$_2$O$_3$ films are strongly affected by the sapphire substrate orientation and thermal-mismatch strain \cite{sakaiTransportPropertiesRatio2015,barazaniPositiveNegativePressure2023}.

The blue shaded curve in Fig. \ref{fig:fig3}(b) illustrates the profile expected if the PM-AFI transition induced the same basal-plane expansion for all grain orientations. It is obtained by uniformly shifting the PM guide curve by a $\Phi$-independent transition jump. The measured AFI data clearly deviate from this reference: the out-of-plane sectors near $\Phi \approx 0^{\circ}$ approach the expected AFI expansion, whereas the in-plane sectors at larger $|\Phi|$ remain closer to the PM-side metric. Thus, similarly to the photoinduced room-temperature response, the deformation appears to be accommodated more freely when the (a$_H$,b$_H$) plane is approximately perpendicular to the substrate plane than when it is closer to parallel to the substrate.

The origin of this orientation dependence can be rationalized by considering the elastic boundary condition imposed by the substrate. After film crystallization at 500 $^{\circ}$C, thermal-expansion mismatch between the film and the sapphire substrate produces an interfacial strain upon cooling \cite{abadiasReviewArticleStress2018, murakamiThermalStrainLead1977, wiederCalculationThermallyInduced1995}. In a textured polycrystalline film, this strain does not act in the same crystallographic direction for all grain families. Figure \ref{fig:fig3}(c) sketches two limiting cases. For $\Phi \rightarrow \pm90^{\circ}$, the (a$_H$,b$_H$) basal plane is parallel to the substrate plane and is therefore directly coupled to the interface. In contrast, for $\Phi \approx 0^{\circ}$, the $c_H$ axis lies in the substrate plane, while the basal plane is approximately perpendicular to the interface; in this geometry, basal-plane strain is expected to arise more indirectly through Poisson coupling.

As a first approximation, the thermal strain accumulated upon cooling can be estimated from the thermal-expansion mismatch between film and substrate\cite{murakamiThermalStrainLead1977, hofmannSolidStatePhysics2015},
\[
\varepsilon_{\mathrm{th}}(T)\propto
\int_{T_{\mathrm{anneal}}}^{T}
\left(
\alpha_{\mathrm{film}}(T')-\alpha_{\mathrm{sub}}(T')
\right)\mathrm{d}T' .
\]

Using literature lattice parameters for V$_2$O$_3$ and (V,Cr)$_2$O$_3$ \cite{mcwhanMetalInsulatorTransitionV1xCrx2O31970}, along with their thermal-expansion coefficients \cite{eckertThermalExpansionCorundum1973,mcwhanMetalInsulatorTransitionV1xCrx2O31970} and that of c-cut sapphire \cite{whiteThermophysicalPropertiesKey1997}, this mismatch can be converted into an expected shift of  $q_{110}$ for the two limiting orientations shown in Fig. \ref{fig:fig3}(c). The representative room-temperature lattice parameters used were $a_H=4.950$ \AA{} and $c_H=14.001$ \AA{} for V$_2$O$_3$, and $a_H=4.999$ \AA{} and $c_H=13.912$ \AA{} for $\mathrm{(V_{0.95}Cr_{0.05})_2O_3}$. For V$_2$O$_3$, the thermal-expansion coefficients were $\alpha_a=20\times10^{-6}$ K$^{-1}$ and $\alpha_c=-8.6\times10^{-6}$ K$^{-1}$ at room temperature, and $\alpha_a=17.9\times10^{-6}$ K$^{-1}$ and $\alpha_c=-0.2\times10^{-6}$ K$^{-1}$ at the annealing temperature. For $\mathrm{(V_{0.95}Cr_{0.05})_2O_3}$, they were $\alpha_a=8.8\times10^{-6}$ K$^{-1}$ and $\alpha_c=3.5\times10^{-6}$ K$^{-1}$ both at room temperature and at the annealing temperature. The in-plane thermal-expansion coefficient of c-cut sapphire was taken as $5.15\times10^{-6}$ K$^{-1}$ at room temperature and $8.36\times10^{-6}$ K$^{-1}$ at the annealing temperature.

The calculated trends are shown in Fig. \ref{fig:fig3}(d). They explain why, already at room temperature before photoexcitation, the azimuthal profiles are expected to be concave. For pure V$_2$O$_3$, the mismatch with sapphire generates a sizeable orientation-dependent basal-plane strain: the orientation in which the basal plane is closer to the substrate plane is expected to show a larger tensile distortion of $a_H$, and therefore a smaller $q_{110}$, than the orientation in which the basal-plane response is mediated through the $c_H$ axis. For (V,Cr)$_2$O$_3$ with 5\% Cr, the thermal-expansion mismatch is much smaller, and the expected orientation dependence is correspondingly weaker.

This picture also provides a framework for interpreting the room-temperature photoinduced data of Fig. \ref{fig:fig2}. At room temperature, the optical excitation is not expected to drive a first-order phase transition, but rather to produce a mainly thermoelastic strain response \cite{thomsenSurfaceGenerationDetection1986,matsudaFundamentalsPicosecondLaser2014,ruelloPhysicalMechanismsCoherent2014, matternConceptsUseCases2023}. The larger photoinduced shift observed in pure V$_2$O$_3$ compared with the Cr-substituted sample is consistent with the larger basal-plane thermal-expansion coefficient of the PM phase relative to the Cr-substituted PI phase. At a fluence of 21 mJ/cm$^{2}$, the estimated temperature rise is $\Delta T \approx 600$ K, which would transiently bring the film from room temperature to approximately 900 K, above the annealing temperature of $\sim 800$ K.

This large transient temperature rise changes the mechanical situation with respect to the laser-off state. Before excitation, the azimuthal strain profile results from thermal-expansion mismatch accumulated during cooling from the annealing temperature to room temperature. After photoexcitation, instead, the film is rapidly heated while the substrate remains comparatively colder on the timescale of the initial structural response. The substrate therefore does not drive the expansion but acts as a colder mechanical boundary that resists the in-plane distortion of the film.


In this transient heating regime, grain families whose (a$_H$,b$_H$) plane is closer to perpendicular to the substrate can expand more freely, whereas those with the basal plane closer to parallel to the substrate experience stronger resistance to the photoinduced deformation. A comparable influence of substrate-induced mechanical constraints on ultrafast strain has been reported in granular FePt thin films, where partial in-plane clamping modifies the transient lattice response \cite{vonreppertUltrafastLaserGenerated2018}. This explains why the photoinduced response is larger near $\Phi \approx 0^{\circ}$ and why the azimuthal curvature can invert with respect to the laser-off profile. In this sense, the room-temperature experiment provides a useful reference: even without crossing the PM-AFI transition, the photoinduced strain is already filtered by the orientation-dependent mechanical coupling to the substrate.

The static measurements therefore establish the mechanical landscape in which the out-of-equilibrium transition takes place. The key question is then how the same orientation-dependent boundary condition also limits the photoinduced AFI$\rightarrow$PM-like transformation\cite{amanoPropagationInsulatortometalTransition2024, lantzUltrafastEvolutionTransient2017, singerNonequilibriumPhasePrecursors2018}. To address this point, we photoexcite the low-temperature AFI phase of pure V$_2$O$_3$ and follow the azimuth-resolved dynamics of the (110)$_H$ reflection.

\begin{figure*}[h!]
\centering
\includegraphics[width=1.0\textwidth]{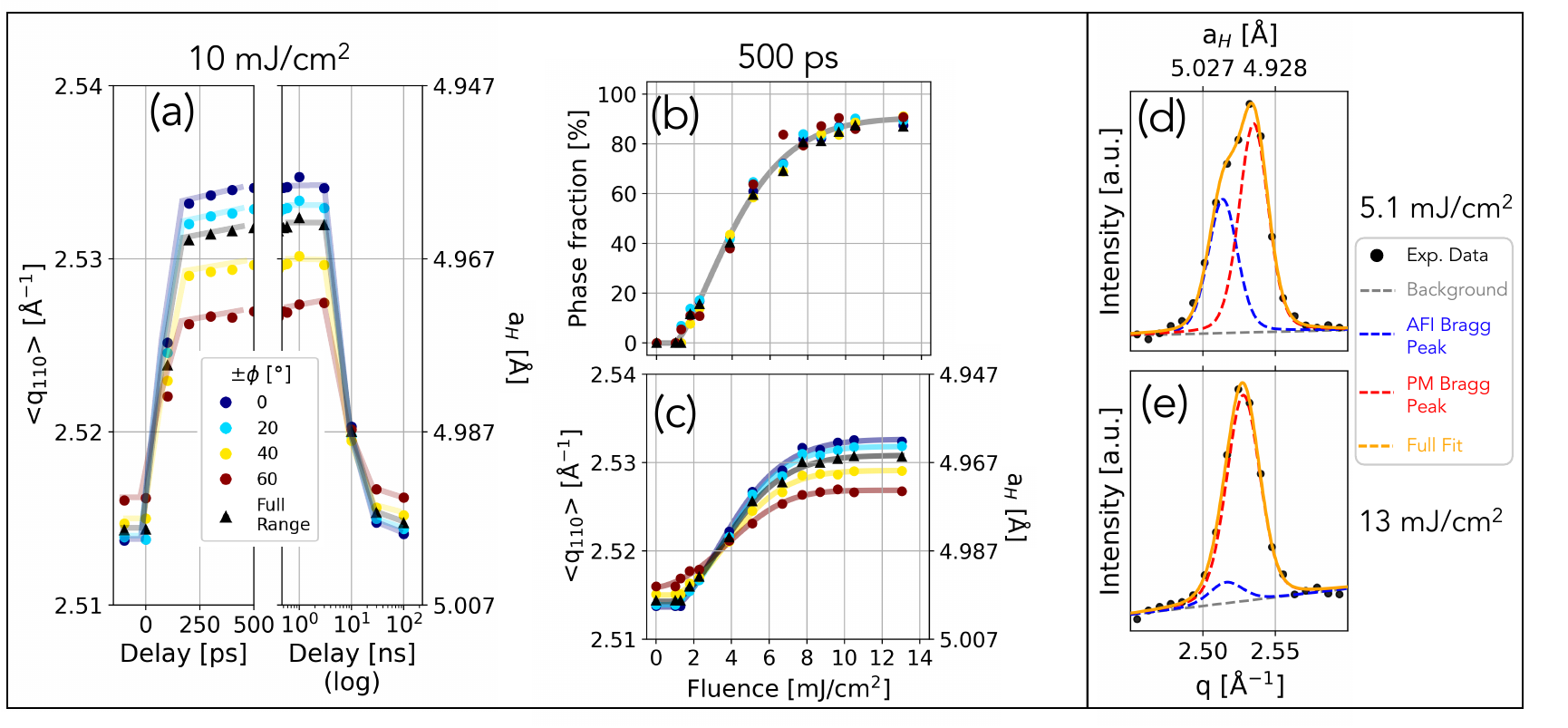}
\caption{\label{fig:fig4}
Out-of-equilibrium response of the $(110)_{H}$ reflection in the AFI phase of pure V$_2$O$_3$ following photoexcitation for different azimuthal angles, measured at T = 80 K. Panels (a)--(c) show azimuthal sectors centered at $|\Phi| = 0^{\circ}$, $20^{\circ}$, $40^{\circ}$, and $60^{\circ}$, together with the fully azimuth-integrated response. (a) Time evolution of $\langle q_{110}\rangle$, and corresponding $a_H$ lattice parameter, after excitation at a fluence of 10 mJ/cm$^{2}$. (b) Photoinduced PM phase fraction at 500 ps as a function of pump fluence. The PM fraction is defined as the area of the photoinduced PM component divided by the total fitted peak area. (c) Average peak position at 500 ps as a function of pump fluence, shown as $\langle q_{110}\rangle$ on the left axis and the corresponding $a_H$ on the right axis. The average position is calculated as the area-weighted mean of the fitted peak positions. (d,e) Representative 500 ps linecuts of the (110)$_H$ reflection for (d) $\Phi \approx 0^{\circ}$ at 5.1 mJ/cm$^{2}$ and (e) $|\Phi| = 60^{\circ}$ at 13 mJ/cm$^{2}$. Experimental data are shown as black circles, the total fit as an orange line, and the individual pseudo-Voigt components associated with the PM-like and AFI-like peaks as red and blue lines, respectively. The comparison between the single- and double-pseudo-Voigt analyses, which give the same azimuthal trends, is provided in Section S3 of the Supporting Information.}
\end{figure*}

Figure \ref{fig:fig4}(a) shows the time evolution of $\langle q_{110}\rangle$ after excitation of the AFI phase at a fluence of 10 mJ/cm$^{2}$. For all azimuthal sectors, photoexcitation shifts the $(110)_{H}$ reflection toward larger $q_{110}$, corresponding to a contraction of the basal plane and therefore to a motion toward the PM-like structural metric. The amplitude of this response is strongly orientation dependent. The largest contraction is observed near $\Phi \approx 0^{\circ}$, while the response is progressively reduced for $|\Phi| = 20^{\circ}$, $40^{\circ}$, and $60^{\circ}$. In contrast, the overall temporal evolution is comparable for the different azimuthal sectors. Thus, the main effect of the crystallite orientation is to set the accessible distortion amplitude, rather than to introduce qualitatively different dynamics.

The laser fluence dependence at 500 ps separates the converted fraction from the average structural distortion. Figure \ref{fig:fig4}(b) shows the PM phase fraction extracted from two-pseudo-Voigt fits of the (110)$_H$ line shape. The fraction of the photoinduced PM-like component increases with fluence, as expected, but no significant azimuthal dependence is observed within the experimental uncertainty. This indicates that, at this delay, the amount of transformed material is mainly governed by the deposited energy.

The situation is different for the average peak position. Figure \ref{fig:fig4}(c) shows the area-weighted mean position of the two fitted components. Although the PM fraction is nearly independent of $\Phi$, the average $\langle q_{110}\rangle$ remains strongly azimuth dependent over the full fluence range. For a given fluence, the displacement from the initial AFI position is largest near $\Phi \approx 0^{\circ}$ and decreases as $|\Phi|$ increases. In other words, fluence controls how much of the material is converted, while the substrate-related orientation determines how far the basal-plane metric can move. This separation is further supported by the fitted position of the photoinduced PM-like component, which remains nearly fluence independent within each azimuthal sector but strongly depends on $\Phi$, as shown in Section S4 of the Supporting Information.

This distinction is illustrated directly by the linecuts in Fig. \ref{fig:fig4}(d,e). At $\Phi \approx 0^{\circ}$ and 5 mJ/cm$^{2}$, the diffraction profile contains two components, indicating coexistence between the initial AFI-like peak and an appearing PM-like peak. At $|\Phi| = 60^{\circ}$ and 13 mJ/cm$^{2}$, the photoinduced PM fraction is larger, but the transformed component remains at a lower $q_{110}$ than in the weakly constrained orientation. As a result, the two cases can display comparable average peak positions while corresponding to different microscopic situations. This shows that the azimuth-resolved line-shape analysis is essential to disentangle the converted fraction from the structural metric reached by each grain family.

Taken together, the static and time-resolved measurements show that the substrate imposes an orientation-dependent mechanical boundary condition on the basal-plane response of the film. Across the thermal PM-AFI transition, this boundary condition suppresses the transition-associated basal-plane expansion in grain families whose basal plane is closer to the substrate plane. After photoexcitation of the AFI phase, the same constraint persists out of equilibrium: weakly constrained grain families reach a larger PM-like basal-plane contraction, while strongly constrained families remain limited to a smaller distortion. The system therefore does not evolve toward a single universal photoinduced lattice state. Instead, fluence primarily controls the converted fraction, whereas interfacial clamping sets the maximum basal-plane distortion accessible within each grain family.

This structural picture also provides a possible framework for understanding why the magnitude and character of the resistivity change at the thermal PM-AFI transition in pseudo-epitaxial V$_2$O$_3$ films depend strongly on substrate orientation, thickness, and growth conditions. Previous studies have reported modified, weakened, or substrate-dependent metal-insulator transition signatures in strained V$_2$O$_3$ films, often correlating these changes with altered $c_H/a_H$ ratios, epitaxial strain, or thermal-expansion mismatch \cite{schulerInfluenceStrainElectronic1997,sakaiTransportPropertiesRatio2015,barazaniPositiveNegativePressure2023,tahaSultanInfluenceSubstrateOrientation2025}. Our azimuth-resolved measurements on granular thin films suggest that such variations in transport properties need not arise only from a change in transformed volume fraction or sample quality: even when a transition is triggered, interfacial clamping can limit the crystallographic distortion reached by selected grain orientations, thereby reducing the effective structural order-parameter jump associated with the insulating or metallic state.

Finally, we note that this azimuth-dependent nonequilibrium structural response is not specific to the V$_2$O$_3$/$c$-cut sapphire system. Additional room-temperature measurements on V$_2$O$_3$ films grown on a different substrate show a similarly anisotropic photoinduced response of the $(110)_H$ reflection: the relative shift $\Delta q_{110}/q_{110,0}$ at $|\Phi| = 60^{\circ}$ is approximately half of that measured near $\Phi \approx 0^{\circ}$, in agreement with the trend observed here, as shown in Section S5 of the Supporting Information. A comparable azimuth-dependent photoinduced response is also observed for the $(202)$ reflection of NiS$_2$, another Mott-insulating thin film, as shown in Section S6 of the Supporting Information. These additional examples indicate that azimuth-dependent ultrafast strain is not governed primarily by the specific substrate or material combination, but by the presence of an interfacial mechanical constraint that converts crystallite orientation into a selector of the nonequilibrium lattice pathway.


\section{Conclusion}

In summary, azimuth- and time-resolved X-ray diffraction shows that thermal-expansion-mismatch strain in textured polycrystalline V$_2$O$_3$ films on $(0001)$ sapphire imposes an orientation-dependent interfacial boundary condition on the basal-plane lattice response. In thermal equilibrium, this constraint appears as a $\Phi$-dependent suppression of the (a$_H$,b$_H$) expansion across the PM$\rightarrow$AFI transition, showing that the structural order-parameter jump is not equivalent for all grain orientations.

Out of equilibrium, the same boundary condition remains active during the photoinduced transition. Even when the excitation fluence is sufficient to generate a PM-like response, grain families that are more strongly coupled to the substrate reach a reduced, clamp-selected lattice metric rather than converging toward a single universal photoinduced metallic structure. The fluence dependence further separates two effects: the deposited energy primarily controls the converted PM-like fraction, whereas the interface limits the maximum basal-plane distortion accessible to each grain family.

This conclusion is not limited to the specific V$_2$O$_3$/sapphire system. The additional measurements reported in the Supporting Information show comparable azimuth-dependent photoinduced strain responses in V$_2$O$_3$ on Si/SiO$_2$ and in NiS$_2$, another Mott-insulating thin film. This point is important for textured or granular films, which may appear structurally three-dimensional when viewed through their diffraction texture or powder-like diffraction rings. Their out-of-equilibrium response, however, is not equivalent to that of a mechanically free powder. Even when many crystallite orientations are present, the film remains coupled to the substrate, and the interfacial boundary condition strongly affects the three-dimensional structural response according to crystallite orientation.

More broadly, these results identify interfacial clamping as a deterministic mechanical selector that co-governs ultrafast switching together with optically generated stress and strain. This perspective complements interface- and drive-based approaches to controlling quantum materials\cite{basovPropertiesDemandQuantum2017} and is directly relevant to Mottronics concepts\cite{tokuraEmergentFunctionsQuantum2017}, where reproducible ultrafast operation in thin-film platforms requires control not only of excitation conditions but also of mechanical boundary conditions.


\begin{acknowledgement}
We thank CNRS for the financial support through the 80$\vert$Prime program (project 'FAST-IA'). E.J, M.L., and C.M. thank the French Agence Nationale de la Recherche for financial support under grant ANR-23-CE30-0027 ('FASTRAIN'). L.C. acknowledges the Région Pays de la Loire for its financial support through the Mott-IA project. We acknowledge ESRF for provision of the synchrotron radiation facilities and use of beamline ID09 under proposal nos. ma5744 and ma5554, and thank R. Garlet, B. Richer and Y. Watier for their help in setting up the cryo-cooling system. We acknowledge MAX IV Laboratory for time on the FemtoMAX beamline under proposal nos. 20250996. Research conducted at MAX IV, a Swedish national user facility, is supported by the Swedish Research Council under contract no. 2018-07152, the Swedish Governmental Agency for Innovation Systems under contract no. 2018-04969 and Formas under contract no. 2019-02496.
\end{acknowledgement}


\bibliography{references}

\end{document}


\maketitle

\renewcommand{\thefigure}{S\arabic{figure}}

\clearpage

\section{S1: Azimuthal calibration using LaB$_6$}

The geometrical calibration used for the azimuthal integration was checked using a LaB$_6$ powder calibrant sealed in a capillary. The diffraction image was collected at an X-ray energy of 15 keV with a sample-detector distance of 15 cm. A representative two-dimensional LaB$_6$ diffraction image is shown in Fig. \ref{fig:SI_fig1}(a). The circular powder rings were used to refine the detector geometry, including the direct-beam position, sample-detector distance, and detector rotations with respect to the diffracting point.

\begin{center}
\includegraphics[width=1.0\textwidth]{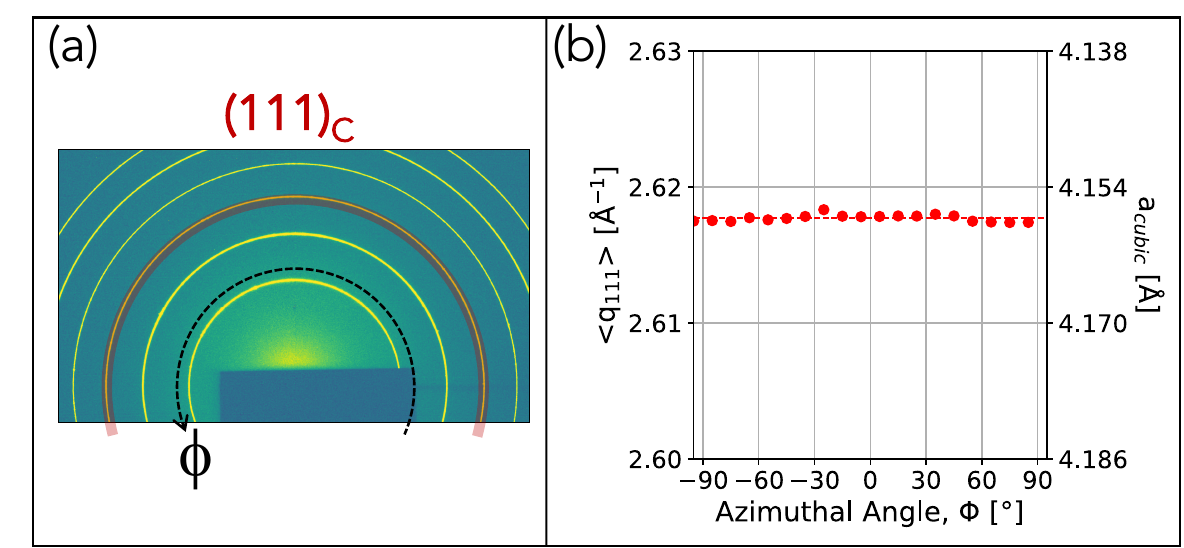}
\end{center}
\refstepcounter{figure}\label{fig:SI_fig1}
\noindent\textbf{Figure \thefigure.} Geometrical calibration using a LaB$_6$ powder calibrant. (a) Representative two-dimensional LaB$_6$ diffraction image collected at 15 keV with a sample-detector distance of 15 cm. The red line highlights the LaB$_6$(111) diffraction ring used to extract the $q_{111}$ values shown in panel (b), while the black arrow indicates the azimuthal angle $\Phi$. (b) Azimuthal dependence of the fitted LaB$_6$(111) peak position, $q_{111}$, extracted from the two-dimensional image in panel (a) using 10$^{\circ}$-wide azimuthal sectors. The corresponding cubic lattice parameter, $a_{cubic}$, is shown on the right axis.

\bigskip

To verify that this geometrical refinement does not introduce artificial azimuthal distortions, the LaB$_6$ image was then integrated over 10$^{\circ}$-wide azimuthal sectors. For each sector, the LaB$_6$(111) reflection was fitted with a pseudo-Voigt profile in the $q = 2.55$-$2.70 \mathrm{\AA^{-1}}$ range. The fitted $q_{111}$ position was converted into the corresponding cubic lattice parameter, $a_{cubic}$, and compared as a function of the azimuthal angle $\Phi$.

As shown in Fig. \ref{fig:SI_fig1}(b), the fitted $q_{111}$ position remains flat over the full azimuthal range. This confirms that the detector geometry was properly refined and that the caking procedure does not generate a measurable artificial azimuthal dependence. 

\section{S2: Thickness dependence of the azimuthal strain}

The thickness dependence of the azimuthal strain was examined by comparing two granular V$_2$O$_3$ thin films with thicknesses of 100 nm and 280 nm. The measurements were performed at 15 keV and analyzed using azimuthal caking of the two-dimensional diffraction images. For each 20$^{\circ}$-wide azimuthal window, the intensity was integrated into a one-dimensional profile and the $(110)_{H}$ reflection was fitted with a pseudo-Voigt function with fixed Lorentzian/Gaussian ratio $\eta = 0.3$. The same fitting procedure was applied to both films so that relative shifts in the extracted $\langle q_{110}\rangle(\Phi)$ values can be attributed to changes in the lattice metric rather than to differences in the analysis.

As shown in Fig. \ref{fig:SI_fig2}, both films display a similar azimuthal dependence of the $(110)_{H}$ peak position. The 100 nm film shows larger $\langle q_{110}\rangle$ values near $\Phi \approx 0^{\circ}$ and lower values at larger $|\Phi|$, corresponding to a larger tensile distortion of the basal plane when the (a$_H$,b$_H$) plane becomes closer to parallel to the substrate. The fully integrated value, shown by the dashed line, lies between the azimuth-resolved extrema, as expected for an average over the textured grain population.

\begin{center}
\includegraphics[width=1.0\textwidth]{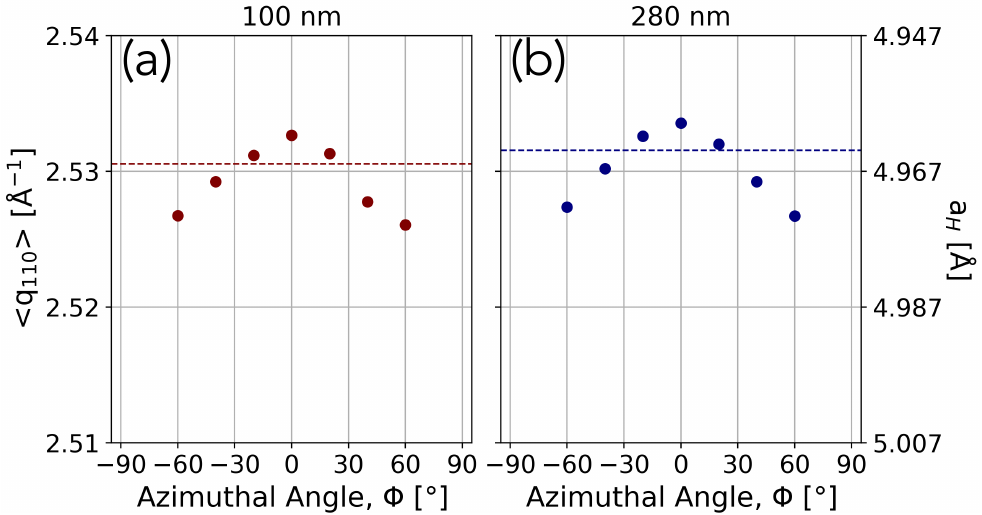}
\end{center}

\refstepcounter{figure}\label{fig:SI_fig2}
\noindent\textbf{Figure \thefigure.} Azimuthal dependence of the V$_2$O$_3$ $(110)_{H}$ peak position, $\langle q_{110}\rangle$, and corresponding $a_H$ lattice parameter for (a) 100 nm and (b) 280 nm granular thin films. Dashed lines indicate the fully integrated values. The small shift observed for the thicker film indicates only weak relaxation of the tensile basal-plane strain.

\bigskip

The 280 nm film preserves the same angular dependence but is shifted only slightly toward higher $\langle q_{110}\rangle$, equivalently smaller $a_H$. This indicates a weak reduction of the tensile basal-plane strain with increasing thickness, but without complete relaxation of the interfacial strain. Even after increasing the film thickness by nearly a factor of three, the orientation-dependent strain remains clearly visible, showing that the thermal-expansion-mismatch strain is only weakly softened over this thickness range \cite{murakamiThermalStrainLead1977}.

\section{S3: Single- and double-pseudo-Voigt analysis}

The fluence-dependent position of the V$_2$O$_3$ $(110)_{H}$ feature was analyzed using two complementary fitting procedures. In the first approach, the full $(110)_{H}$ diffraction feature was described with a single pseudo-Voigt profile with fixed Lorentzian/Gaussian ratio $\eta = 0.3$. This is the same type of analysis used to extract the average peak position in the time-dependent measurements.

In the second approach, the same diffraction feature was fitted with two pseudo-Voigt profiles, also with fixed Lorentzian/Gaussian ratio $\eta = 0.3$, to account for the coexistence between the initial AFI-like component and the photoinduced PM-like component. The position of the initial AFI component was first determined from the laser-off data, where it was allowed to vary freely.

This initial AFI position was then fixed for the fluence series, while the remaining parameters of the two components were left free during the fits. A 5\% area-fraction threshold was applied to avoid overinterpreting weak fitted components. When one of the two components contributed less than 5\% of the total fitted area, this component was set to zero and the line shape was treated as a single-feature response.

This criterion is particularly relevant at large $|\Phi|$, where weak residual components can have intensities comparable to the fitting uncertainty. The average position reported for the double-fit analysis corresponds to the center of mass of the retained fitted components.

As shown in Fig. \ref{fig:SI_fig3}, both fitting approaches lead to the same main conclusion: the fluence-dependent average position of the $(110)_{H}$ feature remains strongly azimuth dependent. The single-pseudo-Voigt analysis gives slightly higher $\langle q_{110}\rangle$ values above approximately 6 mJ/cm$^{2}$, because the remaining AFI component becomes weak and is not explicitly separated from the photoinduced component.

\begin{center}
\includegraphics[width=1.0\textwidth]{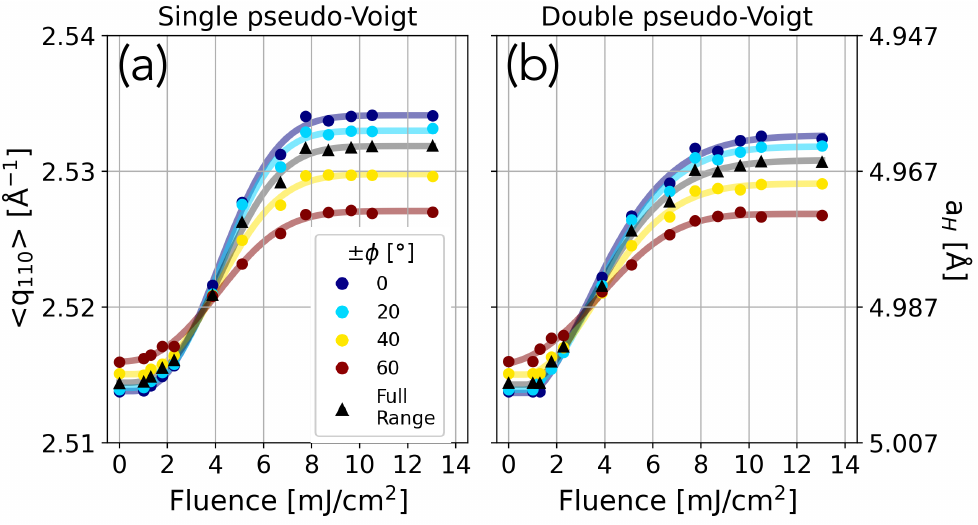}
\end{center}

\refstepcounter{figure}\label{fig:SI_fig3}
\noindent\textbf{Figure \thefigure.} Fluence dependence of the azimuth-resolved V$_2$O$_3$ $(110)_{H}$ response extracted using (a) single and (b) double pseudo-Voigt fits, both with Lorentzian/Gaussian ratio $\eta = 0.3$. Symbols are experimental data and transparent lines are empirical threshold-saturation fits. 

\bigskip

In the double-pseudo-Voigt analysis, this residual AFI contribution is still included in the center of mass when it remains above the 5\% threshold, and therefore shifts the average position slightly toward lower $q_{110}$. The agreement between the two analyses confirms that the azimuthal dependence of the photoinduced response does not result from the particular fitting model used to describe the diffraction feature.

Instead, the extracted trend reflects a robust orientation dependence of the structural response, with the largest displacement toward the PM-like metric occurring near $\Phi \approx 0^{\circ}$ and smaller displacements observed at larger $|\Phi|$.

\section{S4: Azimuthal dependence of the photoinduced strained component}

The double-pseudo-Voigt analysis allows the position of the photoinduced PM-like component to be followed independently from the initial AFI-like component. This is useful because the average peak position depends both on the converted fraction and on the lattice metric reached by the photoinduced phase. To focus on reliable fitted positions, we restrict this analysis to fluences above 4 mJ/cm$^{2}$, where the appearing PM-like fraction exceeds 50

\begin{center}
\includegraphics[width=0.75\textwidth]{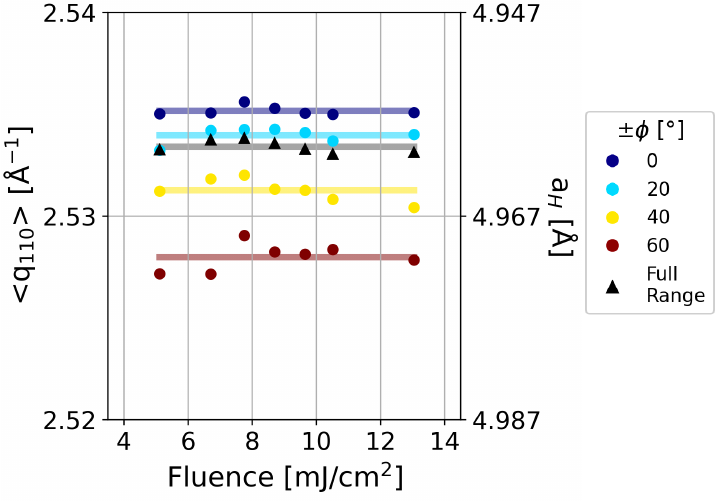}
\end{center}

\refstepcounter{figure}\label{fig:SI_fig4}
\noindent\textbf{Figure \thefigure.} Fluence dependence of the fitted position of the photoinduced PM-like $(110)_{H}$ component for fluences above 4 mJ/cm$^{2}$, where the appearing phase fraction exceeds 50\%. The corresponding $a_H$ lattice parameter is shown on the right axis. Horizontal lines indicate constant fits for each azimuthal sector.

\bigskip

As shown in Fig. \ref{fig:SI_fig4}, the fitted position of the PM-like component is nearly independent of fluence within each azimuthal sector. This is highlighted by the constant fits, indicating that increasing the fluence does not significantly shift the PM-like component toward a different lattice position within a given azimuthal sector.

In contrast, the position of the PM-like component depends strongly on the azimuthal sector. The largest $q_{110}$ value, corresponding to the smallest $a_H$ lattice parameter, is obtained near $\Phi \approx 0^{\circ}$. The fitted $q_{110}$ value then decreases progressively when moving toward larger $|\Phi|$, reaching the smallest value at $|\Phi| = 60^{\circ}$.

\bigskip

\section{S5: Room-temperature photoinduced response of V$_2$O$_3$ on Si/SiO$_2$}

Additional room-temperature pump-probe measurements were performed on a 100 nm granular V$_2$O$_3$ thin film deposited on a Si/SiO$_2$ substrate. The experiment was carried out at the FemtoMAX beamline at MAX IV using 8.7 keV X-rays and a Pilatus 2M detector placed at a sample-detector distance of approximately 11 cm.

The sample was photoexcited at 1500 nm with a fluence of 52 mJ/cm$^{2}$. The laser pulse duration was 50 fs and the X-ray pulse duration was approximately 200 fs. For the analysis shown here, the data were sorted using 1 ps temporal bins to improve the signal-to-noise ratio of the 2D diffraction images. The detector was protected by a Mylar window, which produces diffuse intensity near the center of the Debye rings.

\begin{center}
\includegraphics[width=1.0\textwidth]{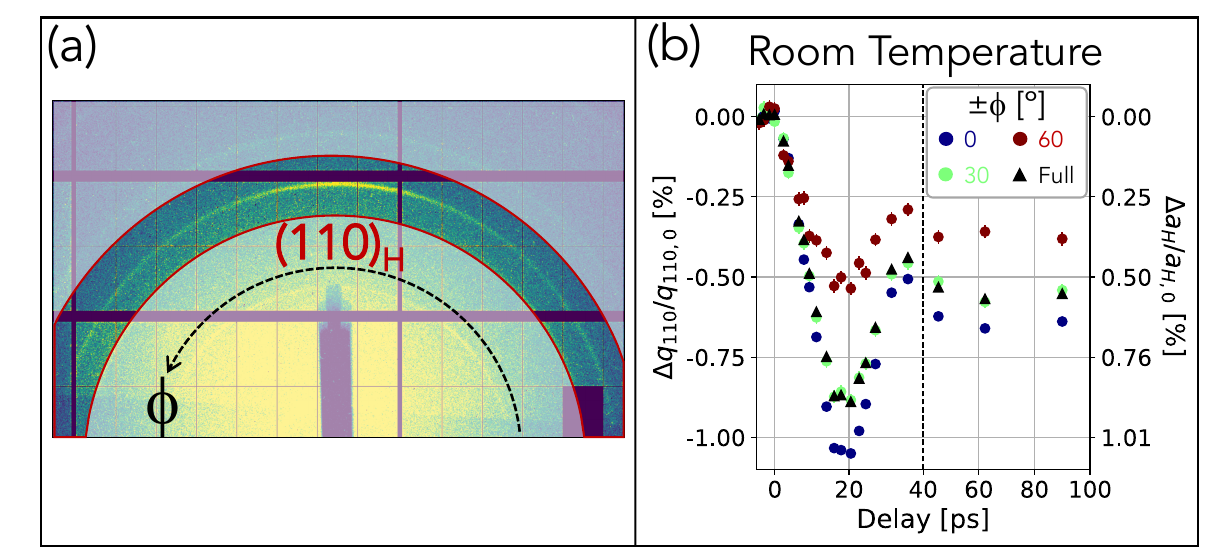}
\end{center}

\refstepcounter{figure}\label{fig:SI_fig5}
\noindent\textbf{Figure \thefigure.} Room-temperature azimuth-resolved photoinduced response of a 100 nm granular V$_2$O$_3$ thin film deposited on Si/SiO$_2$. (a) Representative 2D diffraction image showing the V$_2$O$_3$ $(110)_H$ diffraction ring used for the azimuthal analysis. (b) Time evolution of the relative shift $\Delta q_{110}/q_{110,0}$, and corresponding $\Delta a_H/a_{H,0}$, after photoexcitation. Data are shown for azimuthal sectors centered at $\Phi = 0^{\circ}$, $|\Phi| = 30^{\circ}$, and $|\Phi| = 60^{\circ}$, together with the fully azimuth-integrated response.

The reference position $q_{110,0}$ was defined independently for each azimuthal sector as the average fitted peak position at negative pump-probe delays. The relative shift $\Delta q_{110}/q_{110,0}$ was then calculated with respect to this laser-off reference. The corresponding $\Delta a_H/a_{H,0}$ value was obtained from the inverse relation between $q_{110}$ and the basal-plane lattice parameter. Error bars represent the uncertainty of the fitted peak positions propagated to the relative shifts.

As shown in Fig. \ref{fig:SI_fig5}(b), the photoinduced shift of the $(110)_H$ reflection is strongly azimuth dependent. The maximum relative shift is observed for the sector centered at $\Phi = 0^{\circ}$, whereas the response at $|\Phi| = 60^{\circ}$ is reduced by approximately a factor of two.

This anisotropy closely matches the room-temperature response of V$_2$O$_3$ films on c-cut sapphire discussed in the main text. The observation of a comparable orientation dependence for V$_2$O$_3$ films on a different substrate indicates that the anisotropic nonequilibrium strain response is not governed primarily by the specific substrate material.

Instead, it reflects the presence of an interfacial mechanical boundary condition that constrains the photoinduced lattice expansion according to crystallite orientation. The oscillatory component superimposed on the transient strain response is consistent with coherent acoustic dynamics producing the ultrafast lattice expansion.

\bigskip

\section{S6: Room-temperature photoinduced response of NiS$_2$ on sapphire}

To test whether the azimuth-dependent photoinduced strain response is specific to V$_2$O$_3$, additional pump-probe X-ray diffraction measurements were performed on a NiS$_2$ thin film deposited on sapphire. NiS$_2$ is another Mott-insulating compound, providing a distinct material platform for assessing the generality of the interfacial clamping effect.

The measurements were carried out at room temperature at the ID09 beamline of the ESRF. The sample was photoexcited with 0.5 eV laser pulses at a fluence of 10.5 mJ/cm$^{2}$. The laser pulse duration was approximately 2 ps, while the X-ray pulse duration was approximately 100 ps. The horizontal error bars in Fig. \ref{fig:SI_fig6} represent this X-ray-limited time resolution.

The NiS$_2$ $(202)$ reflection was fitted using a pseudo-Voigt profile on a linear background. The Lorentzian-to-Gaussian contribution was fixed to 0.3. The fitted peak positions were used to extract the relative change $\Delta q_{202}/q_{202,0}$ for azimuthal sectors centered at $\Phi = 0^{\circ}$, $|\Phi| = 20^{\circ}$, $|\Phi| = 40^{\circ}$, $|\Phi| = 60^{\circ}$, and $|\Phi| = 80^{\circ}$, together with the fully azimuth-integrated response.

The corresponding pseudo-cubic lattice response $\Delta a_{\mathrm{pc}}/a_{\mathrm{pc},0}$ was obtained from the cubic relation between the scattering vector and the lattice parameter for the $(202)$ reflection. The reference position $q_{202,0}$ was defined from the laser-off diffraction condition, and the relative shifts were calculated with respect to this reference.

\begin{center}
\includegraphics[width=1\textwidth]{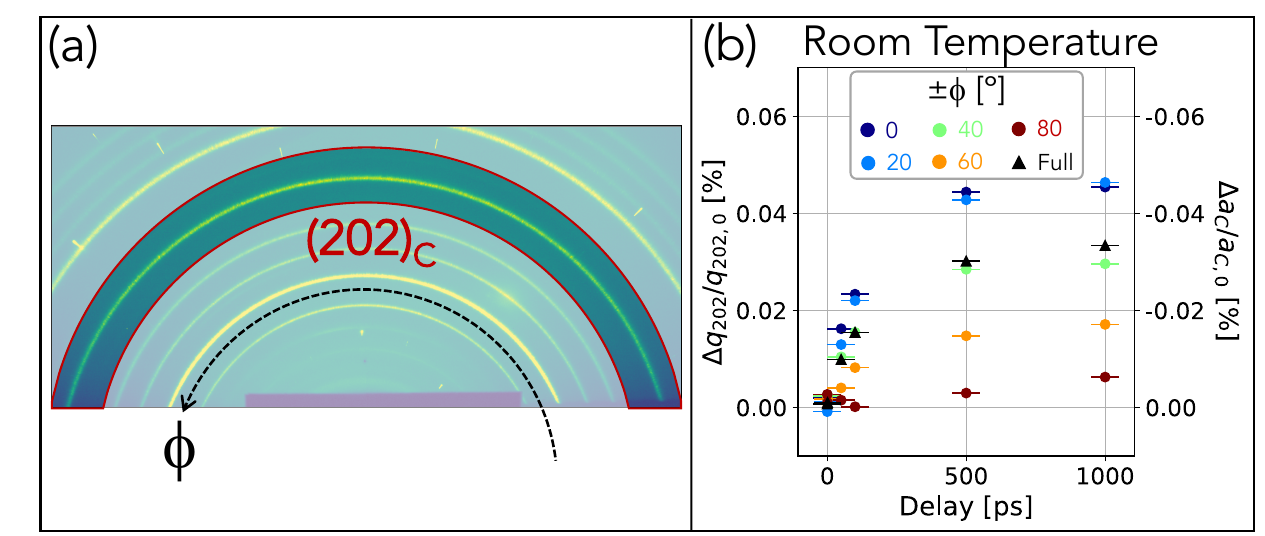}
\end{center}

\refstepcounter{figure}\label{fig:SI_fig6}
\noindent\textbf{Figure \thefigure.} Room-temperature azimuth-resolved photoinduced response of a NiS$_2$ thin film deposited on sapphire. Time evolution of the relative shift $\Delta q_{202}/q_{202,0}$, and corresponding pseudo-cubic lattice response $\Delta a_{\mathrm{pc}}/a_{\mathrm{pc},0}$, after photoexcitation. Data are shown for azimuthal sectors centered at $\Phi = 0^{\circ}$, $|\Phi| = 20^{\circ}$, $|\Phi| = 40^{\circ}$, $|\Phi| = 60^{\circ}$, and $|\Phi| = 80^{\circ}$, together with the fully azimuth-integrated response. Horizontal error bars represent the 100 ps time resolution.

As shown in Fig. \ref{fig:SI_fig6}, the photoinduced shift of the NiS$_2$ $(202)$ reflection is strongly azimuth dependent. The largest response is observed for the sector centered at $\Phi = 0^{\circ}$, whereas the response at $|\Phi| = 60^{\circ}$ is reduced by approximately a factor of two. This behavior closely mirrors the azimuthal trend observed for V$_2$O$_3$ in both the main text and Section S5.

The observation of a comparable anisotropic response in NiS$_2$ shows that the effect is not specific to V$_2$O$_3$. Together with the V$_2$O$_3$/SiO$_2$ result presented in Section S5, these measurements indicate that interfacial clamping of the out-of-equilibrium lattice response is a generic feature of granular thin films, rather than a property of a particular material/substrate combination.

\bibliography{references}